# First-Principles Understanding of Vibrational Energy Transfer in Molecule-Surface Scattering: Both Adiabatic and Nonadiabatic Channels Matter

Gang Meng[1,2] and Bin Jiang[1,2,3]*

[1]State Key Laboratory of Precision and Intelligent Chemistry, University of Science and Technology of China, Hefei, Anhui, 230026, China.

[2]School of Chemistry and Materials Science, Department of Chemical Physics, University of Science and Technology of China, Hefei, Anhui, 230026, China.

[3]Hefei National Laboratory, University of Science and Technology of China, Hefei, 230088, China.

*: Corresponding author: bjiangch@ustc.edu.cn

# ABSTRACT

Energy transfer during molecular collisions on metal surfaces plays a pivotal role in a host of critical interfacial processes. Despite significant efforts, our understanding of relevant energy transfer mechanisms, even in an extensively-studied benchmark like NO scattering from Au(111), remains far from complete. To fully disentangle different energy transfer channels, we develop a first-principles nonadiabatic dynamical model that incorporates explicitly all degrees of freedom and the interfacial electron transfer. Our simulations reproduce, for the first time, most experimental observations on vibrational relaxation and excitation of NO molecules under varying initial conditions and clearly elaborate the respective adiabatic and nonadiabatic contributions. This model identifies the direct adiabatic vibration-to-translation coupling as the dominant role of translation, while excluding significant translation-to-electron nonadiabatic coupling. Furthermore, the observed steric effect varying with the initial vibrational state is understood by the change of orientational dependence of the metal-to-molecule electron transfer. These new insights highlight the importance of treating both adiabatic and nonadiabatic energy transfer pathways on an equal footing, offering significant implications for modeling energy transfer processes in more complex systems, such as plasmonic photocatalysis.

**INTRODUCTION**

Chemical reactions at metal surfaces have profound implications to interfacial applications like heterogeneous catalysis, crystal growth, and etching. Vibrational energy transfer (VET) in molecule-metal surface encounters is highly relevant to surface reactions, as molecular vibration corresponds exactly to the motion of bond breaking/forming(*1-4*). Vibrational energy can be transferred to molecular translation, rotation and surface phonons, an adiabatic process that can be described within the Born-Oppenheimer approximation (BOA)(*5, 6*). Beyond that, VET on metal surfaces is more complicated due to the nonadiabatic energy exchange between molecular vibration and electron-hole pairs (EHPs) of the metal, representing a breakdown of BOA(*7-9*). There has been ample evidence of this nonadiabatic VET channel, causing vibrational excitation/deexcitation, electron emission, or short vibrational lifetimes(*10-15*). Such nonadiabatic processes are also closely related to plasmonic catalysis, where hot electrons generated by plasmon decay may strongly couple to molecular vibration, resulting in vibrationally excited molecules that are more prone to dissociation than thermal ones(*16-19*). An in-depth understanding of the VET dynamics at metal surfaces is thus highly desirable.

However, our understanding on different VET channels remains far from complete, due largely to the lack of an accurate theoretical description of both adiabatic and nonadiabatic processes on equal footing. NO scattering from Au(111) represents one of the most important examples in this aspect, attracting extensive experimental and theoretical attention(*8, 12, 13, 15, 20-37*). State-to-state measurements by Wodtke and coworkers have revealed a number of fascinating features, including the multi-quantum vibrational relaxation of NO ($v_i > 0$) and its dependence on the incidence translational energy ($E_i$) and orientation(*13, 25, 27, 29*), the vibrational excitation of NO($v_i = 0$) and its dependence on surface temperature ($T_s$) and $E_i$(*15, 30*). Such abundant experimental data have provided detailed information on how molecular vibration couples to other degrees of freedom (DOFs), serving as valuable benchmarks to stringently test nonadiabatic theories(*20-22, 24, 31, 33, 34, 36-40*).

Despite substantial efforts in the past two decades, no theory has been so far capable of quantitatively reproducing most experimental observations for NO collisions on Au(111)(*1*). The greatest challenge is to describe the nonadiabatic interactions between the molecular states and the metallic continuum in full-dimensionality. To make the simulation tractable, most theoretical studies have relied on mixed quantum-classical (MQC) approaches, which treat the nuclear motion classically while electronic motion quantum mechanically. Their coupling is either addressed perturbatively on top of the ground state, as in molecular dynamics with electronic friction (MDEF)(*20, 33, 41*), or explicitly by considering two local electronic configurations as a result of electron transfer, as in the independent electron surface hopping (IESH)(*21-23, 37*) and broadened classical master equation (BCME) methods(*36, 42*). Fully quantum dissipative(*20, 40*) and semi-classical(*39*) approaches have also been applied to this system, though limited to only two nuclear DOFs. Indeed, only MDEF and IESH models have ever been integrated with realistic full-dimensional (FD) potential energy surfaces (PESs) to enable meaningful comparisons with experimental data(*15, 21, 33, 37*). Such comparisons based on empirical PESs parameterized by Roy, Shenvi, and Tully (referred to as RST PESs hereafter) suggested an exclusive electron transfer mediated VET mechanism, which can be characterized by IESH yet not MDEF(*15, 23, 37*). However, both methods failed to capture the translational and steric dependence of vibrational inelasticity of NO($v_i = 3$), even qualitatively(*25, 37, 43*). These failures were attributed to the inaccuracies of the RST PESs, presumably arising from the approximate charge-transfer states and less expressive empirical functions(*25, 37*).

More recently, we developed a more reliable ground state FD-PES by embedded atom neural network (EANN) fitting of thousands of density functional theory (DFT) data, on which Born-Oppenheimer molecular dynamics (BOMD) predicted very efficient adiabatic VET(*32, 35*). MDEF simulations on this PES better described the single-quantum vibrational relaxation, but still largely underestimated the extent of multi-quantum vibrational relaxation(*33*). To go beyond the friction model, we applied a constrained DFT (CDFT) scheme(*44, 45*) to determine charge transfer (diabatic) states of molecules at metal surfaces, which was found to more reasonably describe the electron transfer between CO and NO molecule and metal surfaces(*46, 47*). The combination of CDFT, NN, and IESH methods enables a first-principles description for nonadiabatic effects of molecular scattering from metal surfaces.

In this work, we demonstrate the success of this strategy in NO scattering from Au(111) using newly developed EANN-based first-principles charge-transfer state PESs. By choosing an appropriate density functional, IESH simulations on new PESs achieved unprecedented agreement with a wide range of experimental observations of NO on Au(111). Our results emphasize the necessity of properly characterizing both adiabatic and nonadiabatic energy transfer channels, towards a quantitative theoretical description for the VET dynamics of molecules at metal surfaces. The remaining discrepancies with experimental data are also discussed.

**RESULTS**

**Comparison of full-dimensional and two-dimensional dynamics**

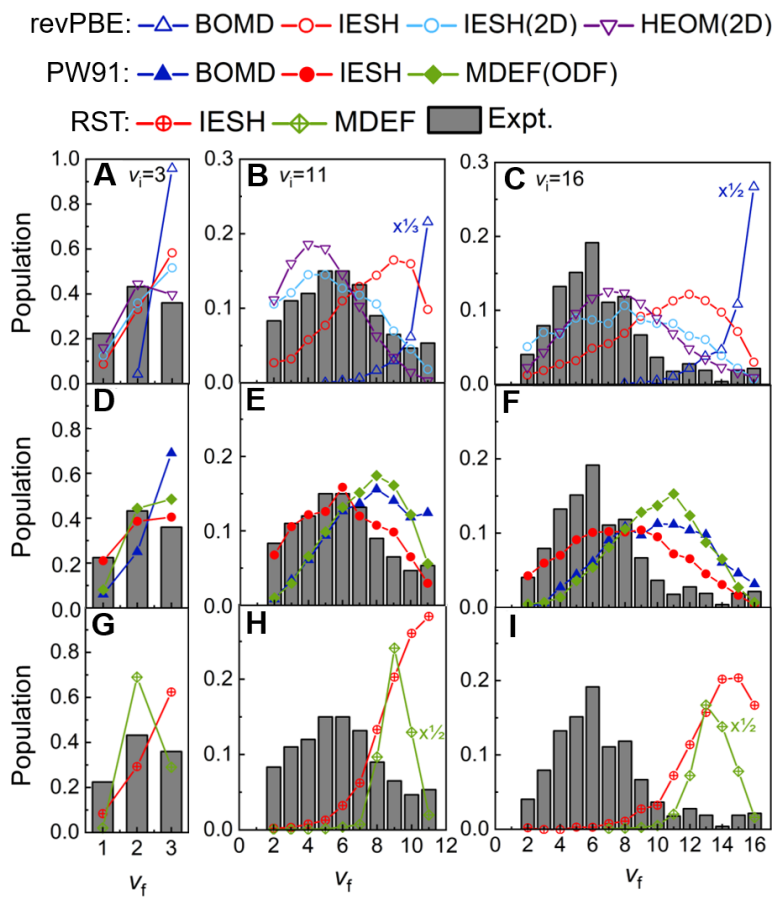

**Fig. 1. Final vibrational state distributions with various initial conditions.**

Comparison of experimental final vibrational state distributions(*27*) of vibrationally excited NO scattering from Au(111) with BOMD and IESH results on the revPBE-based FD-PESs, as well as IESH and HEOM results on the revPBE-based two-dimensional (2D)PESs from Ref. (*40*), for (**A**) $v_i$ = 3, $E_i$ = 1.08 eV, (**B**) $v_i$ = 11, $E_i$ = 0.95 eV and (**C**) $v_i$ = 16, $E_i$ = 0.52 eV. (**D**)-(**F**) Similar to (**A**)-(**C**), but for BOMD, IESH and MDEF(ODF)(*33*) results on the PW91-based FD-PESs. (**G**)-(**I**) Similar to (**A**)-(**C**), but for IESH and MDEF results on the RST FD-PESs from Ref. (*37*). Note that results are extracted from trajectories with single-bounce and trajectories whose final vibrational states excess the experimental ranges are excluded. Results obtained from all trajectories are shown in Fig. S5.

In Fig. 1, vibrational state distributions for scattered NO obtained by various levels of theory, for initial states of $v_i$=3, 11, and 16, are compared with experiments(*27, 37*). For fair comparison, unless stated elsewhere, our new results are analyzed based on single-bounce trajectories, consistent with prior theoretical work and the direct scattering nature of this process as observed in experiments(*13*). For clarity, our results based on all trajectories are shown in Fig. S5 in the Supplementary Materials (SM) and no significant differences are found.

Figs. 1(A-C) show theoretical results using the same revPBE functional, containing several key characteristics. First, in both 2D models, IESH and a fully quantum method, namely hierarchical equations of motion (HEOM)(*40, 48*), yield similar results, although the latter predicts slightly stronger vibrational inelasticity than the former in the case of NO($v_i$ = 3). The good agreement between them largely validates the reliability of IESH. On the other hand, the FD-IESH results are however quite different from their 2D counterparts, manifesting much hotter vibrational state distributions for NO($v_i$= 11 and 16). Indeed, these 2D results are all based on 2D

empirical PESs fitted to limited number of CDFT points reported in Ref. (*46*), where the NO molecule lies perpendicularly on the hcp site of Au(111) with the N atom pointing down. Electron transfer between the molecule and the surface more easily undergoes in this geometry compared to other sites and orientations (see Fig. S6). Apparently, the effects of molecular orientation, surface site, and surface phonons are all neglected in 2D models. The large discrepancy between FD and 2D results underscores the necessity of employing FD-PESs for accurate theoretical modeling. While 2D models may coincidentally align with experimental data, our FD-IESH results remarkably underestimate the degree of vibrational relaxation, suggesting that the revPBE-based PES yields overly repulsive NO/Au(111) interactions. These findings caution against direct comparisons between experimental data and low-dimensional theoretical models, as the latter may fail to capture the true molecule-surface interactions and dynamics.

**Influence of PES on VET dynamics**

To study the influence of the PES on the adiabatic and nonadiabatic VET dynamics, we have constructed another set of FD-PESs based on the PW91 functional, where the ground state PES was reported previously(*32, 33, 35*) and diabatic state PESs are obtained in this work as described in the Materials and Methods section. Figs. 1(D-F) display results obtained on the PW91-based PESs, where MDEF results with orbital-dependent friction (MDEF(ODF)) are taken from Ref. (*33*). As discussed previously(*32, 33, 35*), PW91-based BOMD results predict remarkable vibrational relaxation already, while the MDEF(ODF) model further increases the probability of single-quantum vibrational relaxation (*i.e.* $\Delta v = 1$), but falls short of improving the muti-quantum vibrational relaxation (*i.e.* $\Delta v > 1$)(*33*). Its failure was attributed to the invoked Markovian approximation, which neglects EHP excitations far away from the Fermi level(*33*). Encouragingly, the IESH model, which treats both electron transfer and EHP excitations on equal footing, shows significantly enhanced multi-quantum vibrational energy loss and achieves the best agreement with experimental observations across all initial conditions to date. Not only does the probability of the two-quanta relaxation ($v_i = 3 \rightarrow v_f = 1$) increase remarkably, but the overall vibrational state distributions of NO($v_i$=11 and 16) also shift to lower states, both of which align more closely with the experimental distributions.

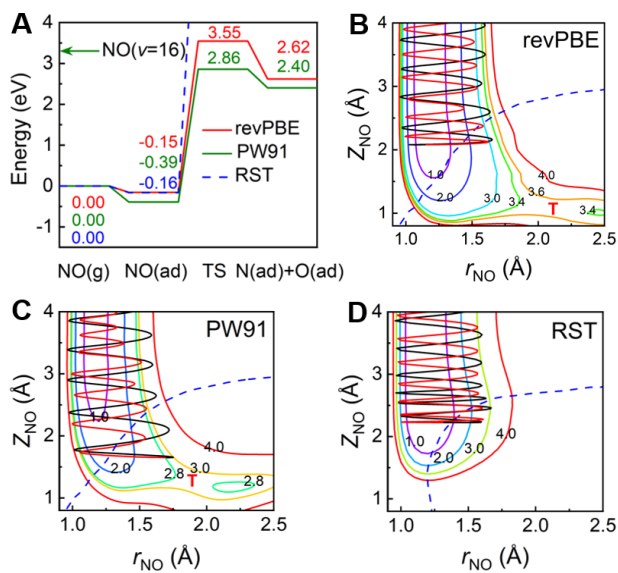

**Fig. 2. Topography of three PESs.** (**A**) Comparison of energies along the minimum energy path of NO dissociation on Au(111) on the ground revPBE, PW91(*35*) and previous RST(*23*) PESs. The green arrow indicates the vibrational energy of NO($v = 16$) in our simulations. Two-dimensional cuts of the (**B**) revPBE, (**C**) PW91 and (**D**) RST ground state PESs as a function of $r_{NO}$ and $Z_{NO}$, with two angles ($\theta$ and $\varphi$, defined in Fig. S1) optimized and other coordinates fixed at the dissociation transition state. The energy is relative to a free NO molecule far from the surface. A representative IESH

trajectory, with the black line representing the incoming component and the red line representing the outgoing component, is projected onto the corresponding PES. The crossing seam between two diabatic states is also shown on the PES. Note that this crossing seam is derived from the orientation that facilitates the most efficient electron transfer, where the NO molecule is perpendicularly placed on the hcp site with the N atom pointing to the surface.

In comparison, PW91-based results represent significant improvements over the revPBE-based and previous RST-based ones(*23, 37*) (Figs. 1(G-I)). On one hand, the adiabatic VET is much more pronounced on the PW91 PES than on the other two PESs (Note that the RST-based BOMD results exhibited minimal vibrational relaxation(*21*), thus not shown here). This adiabatic VET arises from mode softening as the vibrating molecule approaches the dissociation barrier(*32, 49*). As depicted in Fig. 2A, the barrier for NO dissociation on the adiabatic PW91 PES is 0.69 eV lower than that on the revPBE PES, while the RST PES is purely repulsive. In addition, the barrier locates "earlier" on the PW91 PES, featuring a shorter N-O distance, a higher molecule-surface distance, and a wider entrance to the barrier region (see Figs. 2(B-C) and Table S1). These characteristics allow trajectories to access the barrier more readily on the PW91 PES, as demonstrated by an exemplary trajectory of NO($v_i$ = 16) projected onto the 2D PES in Fig. 2. This facilitates significantly more efficient adiabatic VET compared to the other two PESs. On the other hand, in IESH models, the likelihood of electronic excitation is qualitatively linked to the proximity to the crossing seam between neutral and anionic states, where nonadiabatic coupling is strong and electron transfer can explicitly occur, as marked in Fig. 2 by dashed lines. Notably, trajectories on PW91

PESs are more likely to enter this strong coupling region, making electron transfer possible and channeling vibrational energy dissipation into surface EHPs. Indeed, as compared in Tables S2 and S3, the mean vibrational energy losses predicted by PW91-based PESs, due to both adiabatic and nonadiabatic channels, are larger than those predicted by revPBE-based ones. These results reveal that the PES landscape governs simultaneously both adiabatic and nonadiabatic VET channels. This is achieved by modulating the molecular accessibility to the dissociation barrier and state-crossings, which are typically located in similar regions.

Since the PW91-based IESH model appears to describe well both adiabatic and nonadiabatic VET processes, we next extend it to more comprehensive scenarios, including the dependence of vibrational relaxation probability on the incidence energy and initial orientation distribution, the vibration-to-translation coupling, and the vibrational excitation of NO($v_i = 0$). Some of these phenomena have never been adequately described by any previous theoretical models.

**Incidence energy dependence of NO scattering**

Fig. 3 shows the experimentally measured product state branching ratios of NO($v_i = 2$ and 3) as a function of $E_i$. The observed vibrational inelasticity increases monotonically as $E_i$ rises(*12, 25*). PW91-based BOMD results qualitatively capture this $E_i$-dependence, but significantly underestimate absolute ratios. MDEF(ODF) results show some improvement in the vibrationally elastic ($v_f = 3$) and single-quantum relaxed ($v_f = 2$) channels, but have a minimal impact on the two-quanta relaxed ($v_f = 1$) channel.

Impressively, IESH results now well match nearly all branching ratios in the entire range of $E_i$, demonstrating unprecedented agreement with experimental data. It is easily understood that the higher incidence energy allows the NO molecule to get closer to the barrier and state-crossing point, thereby amplifying both adiabatic and nonadiabatic vibrational energy losses. In contrast, previous RST-based IESH results predict a more pronounced vibrational relaxation at lower $E_i$ and thus much weaker $E_i$-dependence of the vibrational inelasticity, as a result of the excessive softness of the RST PES.

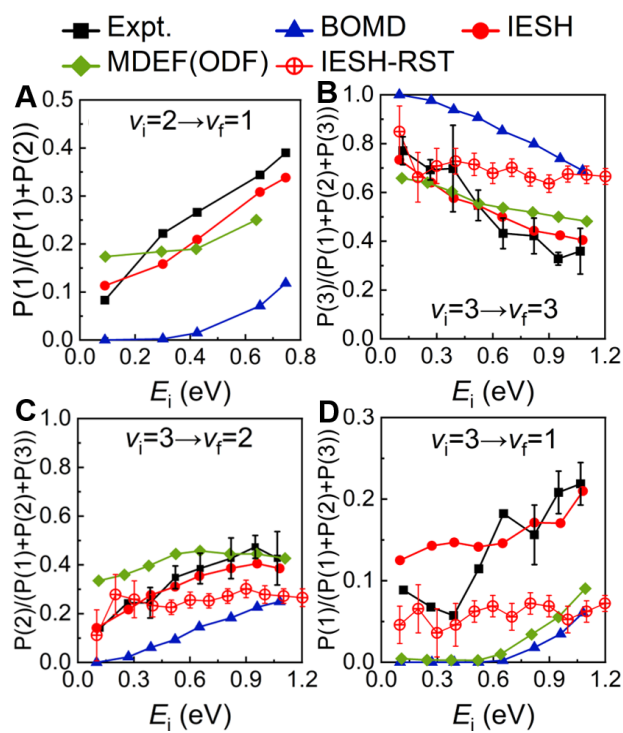

**Fig. 3. Dependence of final vibrational state distributions on $E_i$.** Experimental branching ratios(*12, 25*) as a function of incident translational energy $E_i$, in comparison with BOMD, IESH and MDEF(ODF)(*33*) based on the PW91 PES and IESH using the RST PES(*25*), for (**A**) $v_i = 2$ and (**B**)-(**D**) $v_i = 3$. Note that although our calculations predict minor populations for $v_f = 0$ or $v_f > v_i$, they have not been measured in experiments, which will thus not be discussed here. Only single-bounce trajectories are included in all theoretical models including the referenced MDEF(ODF) and IESH-RST. Results obtained from all trajectories are shown in Fig. S7.

**Orientational dependence of NO scattering**

An intriguing yet not fully understood experimental observation is the orientational dependence of vibrational inelasticity for the scattering of NO($v_i = 3$), which gradually diminishes as the molecule becomes more vibrationally excited(*27, 29*). In experiment, the vibrationally excited NO molecule can be initially manipulated to prefer either an N-first or an O-first orientational distribution before colliding on the surface, as illustrated in Fig. 4A. Fig. 4B shows that the former leads to obviously stronger vibrational relaxation than the latter during NO($v_i = 3$) scattering. However, as the initial vibrational state increases, the orientational dependence largely weakens at NO($v_i = 11$) (Fig. 4C) and nearly vanishes at NO($v_i = 16$) (Fig. 4D). Our PW91-based IESH results, for the first time, properly capture this steric effect and its variation with the initial vibrational state. In Fig. S8, we further quantify the adiabatic and nonadiabatic contributions to the steric effect by comparing BOMD and IESH results with experimental data for NO($v_i = 3$). We assess the degree of vibrational relaxation using the average vibrational quantum number of scattered NO molecules, denoted as $<v_f>$, and quantify the steric effect as the difference of $<v_f>$ between the two initial orientations, namely $<v_f(\text{O-first})>-<v_f(\text{N-first})>$. A larger difference indicates a stronger steric effect. The BOMD prediction of this difference is ~0.13, suggesting a minor orientational dependence due to the adiabatic VET channel. While the IESH prediction increases this value to ~0.21, in close accord with the experimental value of ~0.23. These results demonstrate that both adiabatic and nonadiabatic channels contribute to the steric dependence of VET.

To elucidate the underlying mechanism, we first check the polar angle distributions at the impact point, at which the initially N-first and O-first oriented molecules turn around (see Fig. S9). While some O-first oriented NO molecules reorient towards the N-first orientation, the polar angle distributions in the two cases remain distinct, which is the prerequisite for the stereodynamics. However, these distributions are insensitive to the initial vibrational state, failing to explain the reduced steric effect at higher vibrational states. We then resort to searching the crossings between two diabatic state PESs, plotted as a function of $Z_{NO}$ in Fig. S9, with the polar angle $\theta$ fixed at 68° and 112° — values that correspond to the mean angles ($<\theta>$) for the N-first and O-first distributions, respectively. At $r_{NO}$ = 1.32 Å, approximately the outer classical turning point of NO($v$ = 3), the crossing point of two diabatic states shifts to higher energies as $\theta$ increases. This indicates that N-first oriented molecules would more likely undergo charge transfer than O-first ones and experience more efficient nonadiabatic VET, consistent with their stronger vibrational inelasticity upon collisions. By contrast, when $r_{NO}$ elongates to the turning point of NO($v$ = 16), *i.e.* 1.59 Å, the crossings of two diabatic states become very similar in both orientations. This is in accord with the observed much weaker steric effects for high initial vibrational states. It is worth noting that previous RST-based IESH results turn out to predict no steric effect at all for NO($v_i$ = 3) scattering, mainly due to the too strong dynamical steering (*21, 43, 50*) (see Fig. 4 of Ref. (*21*) and Fig. S9, which presents similar polar angle distributions at the impact point from different initial orientations).

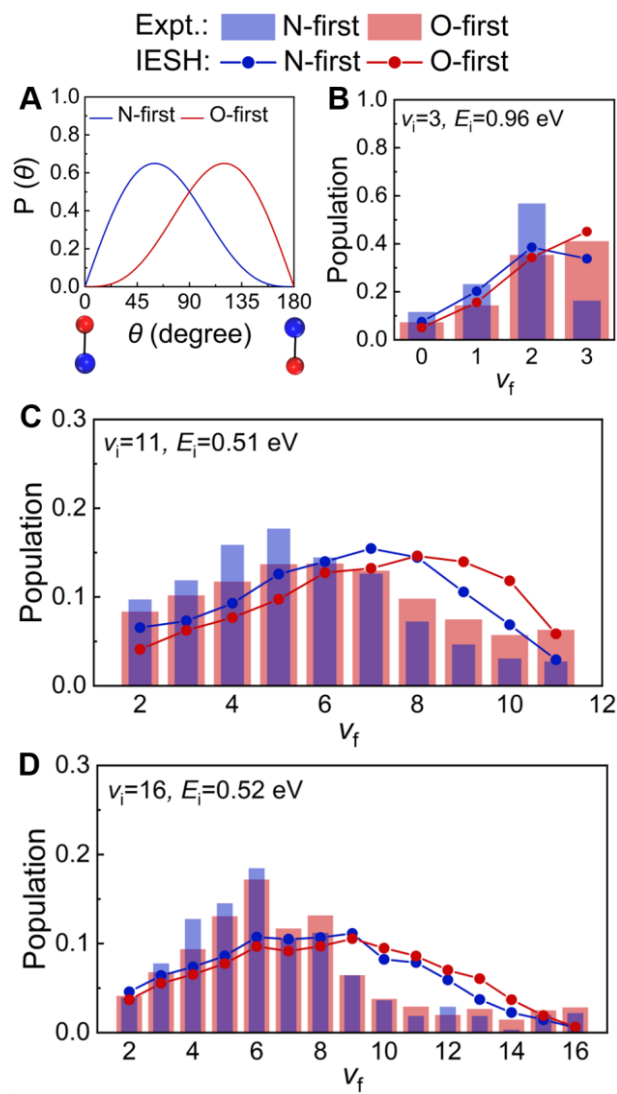

**Fig. 4. Dependence of final vibrational state distributions on initial orientation.** (**A**) Initial polar angle distributions for N-first and O-first orientations of the NO molecule. Experimental final vibrational state distributions(*29*) of NO (**B**) ($v_i = 3$, $E_i = 0.96$ eV), (**C**) ($v_i = 11$, $E_i = 0.51$ eV) and (**D**) ($v_i = 16$, $E_i = 0.52$ eV) scattering from Au(111) with N-first and O-first orientations, in comparison with IESH results on the PW91 PES. Note that only single-bounce trajectories are included, and trajectories with final vibrational states beyond the experimental ranges are not included.

**Vibration-to-translation coupling**

Dynamical results discussed have so far primarily focused on the energy exchange between molecular vibration and surface EHPs (V-EHP). However, the energy transfer process is more complex involving other DOFs, like molecular translation and rotation.

Experimental evidence suggests that the translational motion is not merely a spectator during the electronically non-adiabatic VET process, but its precise role remains elusive. For example, as shown in Fig. 5A, the mean final translational energy ($<E_T>$) is incrementally higher in the vibrationally de-excited channels than the elastic channel(*26, 28*). Experimentalists have proposed two possible mechanisms: a direct mechanical coupling between vibration and translation (V-T), and/or an EHP-mediated coupling, where molecular vibration and translation do not directly couple to each other but both couple to EHPs (V-EHP and EHP-T).(*26, 28*) However, no theory has yet been able to determine which mechanism dominates.

Fig. 5 compares the calculated and experimental $<E_T>$ of scattered NO molecules, as a function of final rotational energy ($E_R$) for three product channels ($v_i = 3 \rightarrow v_f = 3$, 2, 1), at $E_i = 0.98$ eV. The experimentally observed $<E_T>$ decreases linearly with increasing $E_R$ for all three channels, exhibiting an anticorrelation between outbound translation and rotational excitation (T-R). In addition, the evidence of V-T energy transfer is that $<E_{T,v_f=3}>$ extrapolated to $E_R = 0$ is lower by ~0.08 eV than $<E_{T,v_f=2}>$ and ~0.12 eV than $<E_{T,v_f=1}>$, smaller than the corresponding vibrational excitation energy, respectively. In comparison, PW91-based IESH results not only correctly predict the anticorrelation between $<E_T>$ and $E_R$, but also reproduce reasonably well the amount of V-T energy transfer, *e.g.*, ~0.09 eV for $<E_{T,v_f=2}> - <E_{T,v_f=3}>$ and ~0.15 eV for $<E_{T,v_f=1}> - <E_{T,v_f=3}>$, although the slope of d$<E_T>$/d$E_R$ is more negative, implying the anisotropy of the entrance of the PES is less well described.

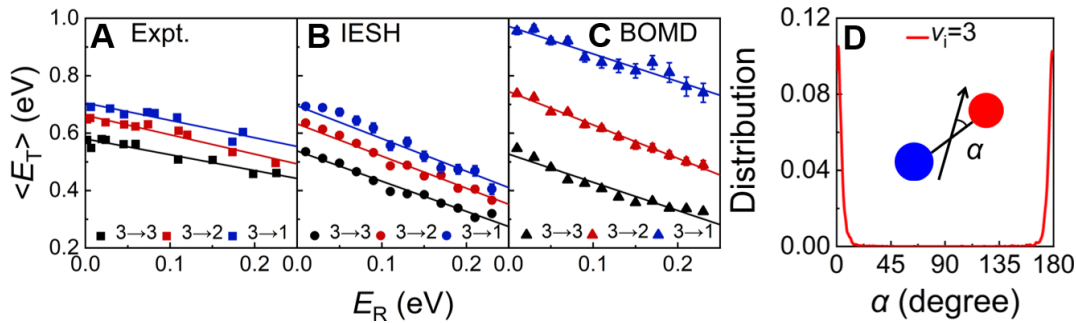

**Fig. 5. Direct vibration-to-translation coupling.** (**A**) Experimental mean final translational energy(*26*) of the scattered NO as a function of final rotational energy for $v_i = 3 \to v_f = 3, 2, 1$ with $E_i = 0.98$ eV. (**B**)-(**C**) Same as (**A**), but calculated by IESH and BOMD on the PW91 PES. (**D**) Distribution of $\alpha$, for which $\alpha$ is the angle between the nonadiabatic coupling vector and N-O bond when electronic hopping occurs in IESH.

Interestingly, Fig. 5C shows that PW91-based BOMD results also exhibit a qualitatively similar anticorrelation between $<E_T>$ and $E_R$ across three well-separated vibrational channels, suggesting a direct V-T coupling with no need to invoke nonadiabatic effects. Indeed, the differences in $<E_T>$ between adjacent vibrational channels are roughly constant at ~0.22 eV, corresponding approximately to the energy of a single vibrational quantum ($\Delta v=1$). This means that vibrational energy can exclusively convert to the outbound translation because of vibrational deexcitation within BOA. When the nonadiabatic channel is open, part of vibrational energy flows to EHPs, thereby reducing the differences of $<E_T>$ between vibrationally relaxed and elastic channels. This analysis also conforms to the mean energy changes in different DOFs during NO($v_i = 3$) scattering (Table S3). Compared to BOMD, the average translational energy loss is barely changed when electronic DOFs are included via IESH, showing no evidence of energy exchange between translational and electronic DOFs. This further confirms the negligible coupling between translation and EHPs. This feature can be understood by analyzing the direction of the nonadiabatic coupling vector whenever surface hopping occurs in IESH simulations. As shown in Fig. 5D, the angular distribution between the nonadiabatic coupling vector and N-O bond ($\alpha$) is sharply peaked at parallel directions, either 0° or 180°. In such scenarios, the nonadiabatic coupling vector is either parallel or antiparallel to the N-O vibrational motion, resulting in the predominance of V-EHP coupling. These results clearly demonstrate that molecular vibrational energy can transfer to both molecular translation and metallic EHPs during scattering, yet the direct T-EHP coupling is negligible, at least in the conditions considered here.

**Vibrational excitation of NO($v_i = 0$) scattering**

Apart from vibrational relaxation of NO at various initial conditions, vibrational excitation of NO($v_i = 0$) was also observed during its scattering from Au(111)(*15, 30*). In this process, the electronic energy is in turn transferred to molecular vibration. Indeed, this process is relevant to plasmonic catalysis, where nonadiabatic energy transfer from hot electrons into vibrationally excited states facilitates the bond cleavage(*19*). As shown in Fig. 6, our PW91-based IESH results compare well with experimental data on the vibrational excitation probabilities for NO ($v_i = 0 \rightarrow v_f = 1$), in a range of $T_s$ from 473 to 973 K at two representative $E_i$ values. Both results are below the thermal limit, indicating that vibrational excitation occurs in a direct scattering process. Experimental vibrational excitation probabilities can be fitted to an Arrhenius equation on $T_s$. PW91-based IESH results generally follow this Arrhenius relation, except the deviation at $E_i$

= 0. 63 eV, $T_s$ = 473 K. The overall agreement observed here is comparable to that of the earlier RST-based IESH model(*15*).

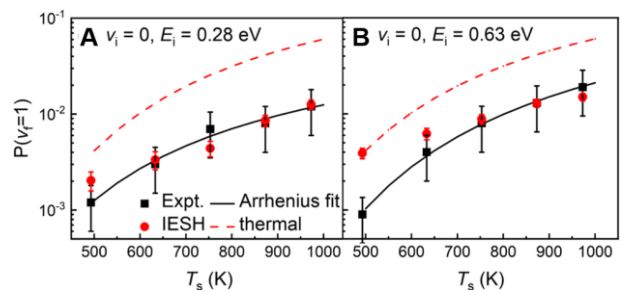

**Fig. 6. Vibrational excitation probability.** Experimental vibrational excitation probability of NO($v_i$ = 0) scattering from Au(111) against $T_s$ for $E_i$ = (**A**) 0.28 eV and (**B**) 0.63 eV(*30*), in comparison with IESH results on the PW91 PES. Arrhenius functions fitted to the experimental points, as well as the thermal limit are also depicted for comparisons.

**DISCUSSION**

Given its significantly improved agreement with experimental results across diverse conditions compared to previous models—unlikely due to error cancellation—the current PW91-based IESH model represents the state-of-the-art in theoretically characterizing molecule-surface scattering processes involving strong nonadiabatic effects. Despite this unprecedented success, some discrepancies persist between the current theory and experiment. One deficiency is the inaccurate prediction for the molecular trapping probability. Fig. 7A shows that the revPBE-based IESH model systematically underestimates the trapping probability of NO($v_i$ = 2) in a range of incidence energies(*51*), particularly at low incident energies, whereas the PW91-based IESH model does the opposite. This discrepancy suggests that neither the revPBE-based nor PW91-based PESs are sufficiently accurate in describing molecular adsorption, with the former being likely too repulsive while the latter too attractive. Another discrepancy between theory and experiment arises in the final rotational state distributions of scattered NO($v_i$ = 3), as illustrated in Figs. 7(B-C). While both models capture the stronger rotational excitation for the O-first than N-first orientation in the vibrationally elastic channel, they improperly predict an artificial rotational rainbow feature for the O-first orientation in the vibrationally inelastic channel and too much rotational excitation, especially the PW91-based IESH model. This is likely tied to the inadequate description of attractiveness, anisotropy and corrugation of the PES, but their influences are coupled and subtle(*52, 53*).

Both deficiencies likely stem from the insufficient accuracy of the PES in the entrance channel described by DFT based on the generalized gradient approximation (GGA). It is possible to take a specific reaction parameter (SRP) approach, which mixes the two density functionals to achieve a more balanced description of the molecule-surface interactions. This strategy has shown some success in accurately predicting experimental dissociation probabilities of $H_2$ on several metal surfaces(*54, 55*). However, its effectiveness in the current system is rather uncertain, because the mix of revPBE and PW91 functionals would unavoidably rise the barrier height compared to that by using PW91 alone and reduce the efficiency of VET. In addition, this strategy will unlikely improve the description for anisotropy and corrugation. More advanced hybrid functionals or doubly hybrid functionals beyond GGA, have potentials for attaining a systematically more accurate description to such systems(*56, 57*), although at the price of much higher computational costs.

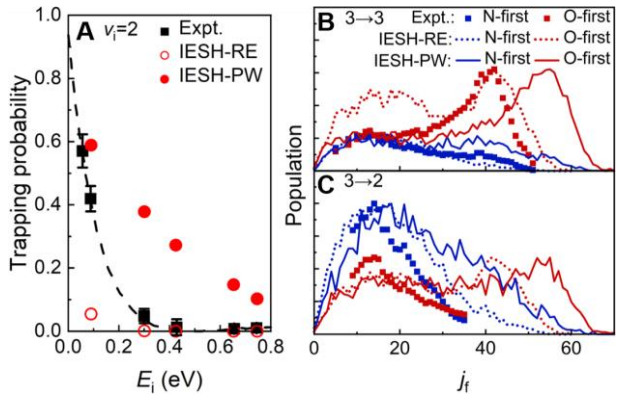

**Fig. 7. Discrepancies with experiments.** (**A**) Trapping probabilities for $v_i = 2$ at a range of incident energies determined by experiments,(*51*) compared to that predicted by IESH on the revPBE (labelled with '-RE') and PW91 (labelled with '-PW') PESs. The black dashed line represents a fit of experimentally determined trapping probabilities. Experimental final rotational state distributions for vibrationally (**B**) elastic ($v_f = 3$) and (**C**) inelastic ($v_f = 2$) scattering of NO($v_i = 3$) from Au(111) with $E_i = 0.96$ eV, for N-first and O-first orientations, compared with IESH results on the revPBE and PW91 PESs. IESH results for both channels are scaled by a factor to match the peaks of the experimental rotational state distributions.

Besides the influence of molecule-surface interactions, the IESH method itself invokes approximations for describing a molecule interacting with an electron bath. First, IESH treats surface electrons independently(*22*), neglecting many-electron effects. Second, like other surface hopping methods, IESH also suffers from the overcoherence issue, although recent studies suggest its impact may be limited(*38, 58*). Nuclear quantum effects could also be significant in scattering processes involving sparse vibrational states(*59*). These limitations may intricately influence the agreement with experiment. Unfortunately, numerically exact quantum methods, such as HEOM, are computationally prohibitive for systems beyond 2D models (*40*). Further research is needed to accurately benchmark the importance of many-electron and overcoherence effects in FD simulations.

Summarizing, this work reports a first-principles model within full-dimensionality to investigate the VET dynamics in NO scattering from Au(111). In this model, two charge-transfer-related diabatic states of NO on Au(111) are calculated by CDFT, which are fitted to diabatic PESs by EANN. These PESs are then used to construct an effective Hamiltonian for the IESH nonadiabatic dynamics simulations, enabling the modelling of explicit electron transfer and EHP excitations on equal footing. Through systematic comparisons of previous 2D models and current FD models based on two different functionals, we find that both adiabatic and nonadiabatic energy transfer dynamics are highly sensitive to the potential energy landscape and model dimensionality. Both adiabatic and nonadiabatic energy transfer processes are found crucial for quantitatively explaining experimental observations. In this regard, our PW91-based IESH model has achieved unprecedented agreement with most experimental data, making a significant step towards a quantitatively accurate understanding of this benchmark system. This CDFT+EANN+IESH scheme scales reasonably with the system size, making it feasible to explore more complex chemical processes involving electron transfer on metal surfaces. Its general success in this model system will hopefully inspire further applications in hot electron-driven chemistry, which could provide first-principles insights into plasmonic photocatalysis(*60, 61*).

**MATERIALS AND METHODS**

In this work, nonadiabatic dynamics simulations for NO scattering from Au(111) are carried out by the IESH method(*22, 62*). To this end, the system is modeled by an impurity level of the molecule interacting with a continuum of metal. Based on a discretized version of the Newns-Anderson model, where the metallic band is approximated by *M* one-electron orbital levels, the many-electron Hamiltonian can be expressed as a sum of one-electron terms(*22*),

$$H_{el}(\mathbf{R}) = U_0(\mathbf{R}) + \sum_{j \in \mathbf{s}(t)} E_j(\mathbf{R}), \quad (1)$$

where $E_j(\mathbf{R})$ is the $j_{th}$ eigenvalue of the following one-electron Hamiltonian,

$$H_{el}^1(\mathbf{R}) = (U_1(\mathbf{R}) - U_0(\mathbf{R}))|a\rangle\langle a| + \sum_{k=1}^{M} \varepsilon_k |k\rangle\langle k| + \sum_{k=1}^{M} V_{ak}(\mathbf{R})(|a\rangle\langle k| + |k\rangle\langle a|). \quad (2)$$

Note that $U_0(\mathbf{R})$ and $U_1(\mathbf{R})$ correspond to the interactions of the neutral molecule (NO) and negative ion (NO⁻) with the surface, $\varepsilon_k$ is the energy of $k_{th}$ metal orbitals, and $V_{ak}(\mathbf{R})$ represents the coupling strength between molecular orbital $|a\rangle$ and metal orbital $|k\rangle$. In this model, the ground state fills the lowest $N_e$ one-electron orbitals ($N_e = M/2$), while excited states are produced by moving one or more electrons to orbitals above the Fermi level, for which these occupied orbitals are indexed in a time-dependent $\mathbf{s}(t)$ vector. Nonadiabatic transitions of independent electrons among one-electron orbitals are then modeled by a modified "fewest switches surface hopping" algorithm(*63*), where the nonadiabatic coupling vector is a one-electron operator that can be computed separately for each electron(*22*). More details about the IESH method are given in the SM.

A key difference between the present and previous IESH models is that $U_0$ and $U_1$

are now determined by CDFT energies with a constrained Bader charge of 0 $e^-$ and -1 $e^-$ onto the NO moiety, respectively. Their couplings are then derived by their energies and the ground state DFT energy ($E_g$) via,

$$V_c = \sqrt{(U_0 - E_g) \times (U_1 - E_g)},  \qquad (3)$$

which is then used to evaluate $V_{ak}$. Specifically, these CDFT energies were calculated with CP2K(*45, 64*), using the same settings as reported in Ref. (*46*). A well-established EANN method(*65, 66*) was employed to learn CDFT energies and forces to yield diabatic PESs, in the same way of representing the ground state PES. This allows us to construct the FD charge-transfer Hamiltonian without introducing any empirical parameters. More details about the CDFT calculations and EANN PESs are given in the SM.

It should be noted that we have constructed PESs based on two density functionals to investigate the influence of the energy landscape on the adiabatic and nonadiabatic VET dynamics. As an extension of Ref. (*46*), the revPBE functional,(*67*) which yields a binding energy of NO with Au(111) in agreement with the experiment value(*68*) (see Table S1), was first used to generate a set of FD-PESs. These PESs allow us to compare the present FD results with previous 2D ones(*40*), which only include the N-O bond length ($r_{NO}$) and the molecular height on the surface ($Z_{NO}$), using the same density functional. However, as discussed above, the revPBE-based PESs appear too repulsive to enable efficient VET. Since a more attractive ground state PES using the PW91 functional is available(*69*), which has already proven to predict reasonably large vibrational energy loss via the adiabatic channel,(*32, 35*) we complement two PW91-based diabatic PESs. Interestingly, as shown in Fig. S2, although individual potential energy curves computed by revPBE and PW91 are rather different, their energy differences between the diabatic state and the ground state, namely ($U_0$-$E_g$) or ($U_1$-$E_g$), are very similar, regardless of the molecular geometry. This is very encouraging and allows us to avoid recomputing all CDFT energies and forces with PW91. Specifically, we added ($U_0$-$E_g$) and ($U_1$-$E_g$) obtained by revPBE-based PESs to the PW91-based ground state, yielding the PW91-based diabatic PESs. More details on the validation of this treatment and the resultant PESs can be found in the SM. These PW91-based PESs were subsequently applied in IESH simulations and compared with a variety of experiments.

**Acknowledgement:** This work is supported by the Strategic Priority Research Program of the Chinese Academy of Sciences (XDB0450101), Innovation Program for Quantum Science and Technology (2021ZD0303301), National Natural Science Foundation of China (22325304 and 22221003). We appreciate the high-performance computing resources provided by the supercomputing center of USTC, Hefei Advanced Computing Center, and Beijing PARATERA Tech CO., Ltd. We sincerely thank Profs. Alec Wodtke, Tim Schäfer, Hua Guo, Reinhard Maurer, and Wenjie Dou, for their insightful discussions.

**Supplementary materials:** Computational details for DFT calculations, neural network potentials, and IESH calculations, along with some additional results in support of the conclusion.

**Data availability:** The reported EANN PESs, DFT, and CDFT datasets are available upon reasonable request to the authors. The used RST PES is downloaded from https://github.com/chinmaypradhan139/IESH-NO_scattering_on_Au/.

# Supplementary Materials for

# First-Principles Understanding of Vibrational Energy Transfer in Molecule-Metal Scattering: Both Adiabatic and Nonadiabatic Channels Matter

Gang Meng[1,2] and Bin Jiang[1,2,3*]

[1]State Key Laboratory of Precision and Intelligent Chemistry, University of Science and Technology of China, Hefei, Anhui, 230026, China.

[2]School of Chemistry and Materials Science, Department of Chemical Physics, University of Science and Technology of China, Hefei, Anhui, 230026, China.

[3]Hefei National Laboratory, University of Science and Technology of China, Hefei, 230088, China.

*: Corresponding authors: bjiangch@ustc.edu.cn

# Table of Contents

# I. Computational Details

## I.A Density functional theory calculations

### I.A1 Conventional density functional theory

To construct a reliable effective Hamiltonian of NO interactive with Au(111) for subsequent nonadiabatic dynamics simulations, we need to calculate both the ground state and the two (quasi)-diabatic states relevant to electron transfer from first-principles. The conventional Khon-Sham density functional theory (DFT) was used to compute ground state energies and forces. In practice, Kohn−Sham orbitals were expanded into mixed Gaussian plane-wave basis in CP2K(*1*) and the ionic cores were approximated by the Goedecker-Teter-Hutter (GTH) pseudopotentials(*2*). Valence electrons were expanded with the DZVP-MOLOPT-SR-GTH basis set for all elements. Since NO is an open shell molecule, all calculations were spin polarized. The surface was modelled by a 6 × 6 supercell with four metal layers. The top two layers of Au atoms were moveable and a vacuum space of 25 Å in the *Z* direction was used to separate the periodic slabs. A Fermi−Dirac smearing method was used with electronic temperature of 1200 K to maintain the fractional occupation of states near the Fermi energy and to speedup self-consistent field convergence. As mentioned in the main text, most calculations in this work have been done using the revPBE functional(*3*), except in some test calculations discussed below, where the PW91 functional was used.

### I.A2 Constrained density functional theory

The constrained DFT (CDFT) method was used to determine the (quasi)-diabatic states of neutral NO molecule and NO anion on the Au(111) surface. According to Wu and Van Voorhis(*4*), in CDFT, the energy functional is defined by modifying the traditional DFT functional, $E[\rho]$, with a Lagrange multiplier ($V$) dependent on the charge constraint,

$$W[\rho,V] = E[\rho]+V(\int w_c(\mathbf{r})\rho(\mathbf{r})d\mathbf{r} - N_c). \tag{S1}$$

Here, $w_c(\mathbf{r})$ is a spatial weight function of the charge density $\rho(\mathbf{r})$, $N_c$ is the target value of the charge constraint. The energy of the charge-constrained state is obtained by self-consistently minimizing $W$ with respect to $\rho$ and maximizing $W$ with respect to $V$,

$$E_{\mathrm{CDFT}} = \min_{\rho} \max_{V} [W[\rho,V]]. \tag{S2}$$

Here, the neutral and the anionic states of NO/Au(111) correspond to 0 $e^-$ or -1 $e^-$ net Bader charge localized in the NO molecule, respectively. The CDFT diabatic state energies were obtained using the CDFT module implemented in CP2K(*5*)—a general way to calculate charge-constrained states for molecule-metal systems(*6*). Compared to the electric field method(*7*), diabatic states defined by CDFT are well-behaved and smooth in the entire dynamically relevant configuration space, enabling constructions of full-dimensional diabatic potential energy surfaces (PES) with accuracy at first-principles level. For the current NO/Au(111) system, the convergence criterion for the charge constraint was set to $5\times10^{-3}$ $e^-$. Other DFT setups were kept consistent in ground state DFT and CDFT calculations.

**I.B Neural network potential energy surfaces**

**I.B1 The revPBE based PESs**

An embedded atom neural network (EANN)(*8, 9*) approach was employed to construct high-dimensional PESs for all ground, diabatic neutral, and anionic states. In this approach, the total energy of the system is decomposed as the sum of atomic energies, each of which is represented by an atomic neural network,

$$E = \sum_{i=1}^{N} E_i = \sum_{i=1}^{N} NN_i(\rho^i),  \tag{S3}$$

where $\rho^i$ are embedded atom density (EAD) features regarded as the input to each atomic neural network that can be simply expressed by the square of the linear combination of Gaussian-type orbitals (GTOs) located at neighbor atoms,

$$\rho^i = \sum_{l_x,l_y,l_z}^{l_x+l_y+l_z=L} \frac{L!}{l_x!l_y!l_z!} (\sum_{j \neq i}^{N_{cut}} c_j^{Z_j} \varphi_{l_xl_yl_z}^{\alpha,r_s}(\mathbf{r}_{ij}) f_c(\mathbf{r}_{ij}))^2.  \tag{S4}$$

In Eq. (S4), $N_{cut}$ is the total number of neighbor atoms within a cutoff radius ($r_c$), $c_j^{Z_j}$ is the element-dependent orbital coefficient which can be optimized together with NNs parameters, $f_c(\mathbf{r}_{ij})$ is a cutoff function to ensure that the contribution of each neighbor atom decays smoothly to zero at $r_c$. The GTO, $\varphi_{l_xl_yl_z}^{\alpha,r_s}(\mathbf{r}_{ij})$, is defined by,

$$\varphi_{l_xl_yl_z}^{\alpha,r_s}(\mathbf{r}_{ij}) = x^{l_x} y^{l_y} z^{l_z} exp(-\alpha|r_{ij} - r_s|^2),  \tag{S5}$$

where $\mathbf{r}_{ij} = (x, y, z)$ is the Cartesian coordinates of the embedded atom *i* relative to a neighbor atom *j*, $r_{ij}$ is its length, $\alpha$ and $r_s$ are parameters that determine radial distributions of GTO, $l_x + l_y + l_z = L$ specifies the orbital angular momentum (*L*).

Data points used to construct the revPBE-based ground, neutral and anionic PESs were generated by two procedures. First, approximately 500 configurations for the NO/Au(111) system were randomly generated and computed to construct an initial ground state PES. The rest 2000 configurations were iteratively selected from quasi-classical trajectories on the iteratively refined ground state PES via both geometric and energetic criterions until the dynamical results converged. Specifically, we calculated the generalized Euclidean distances (GED) between a new configuration and existing data points. A configuration was then selected for the energy criterion test only if all GED values exceeded 1.5 Å. Here, the GED between the $m$th and $n$th configurations is defined as in Ref. (*10*),

$$d_{mn} = \sqrt{(\mathbf{x}^m - \mathbf{x}^n)^2}, \tag{S6}$$

where $\mathbf{x}^m$ and $\mathbf{x}^n$ are the molecular position vectors (for O and N) of the two configurations shifted within the same unit cell. The energetic criterion was defined by the difference of outputs of two neural network fitted with different initial parameters,

$$F(\mathbf{R}) = |y_1(\mathbf{R}) - y_2(\mathbf{R})|, \tag{S7}$$

where $\mathbf{R}$ is the Cartesian coordinate vector of a configuration of the NO/Au(111) system, $y_1$ and $y_2$ are the two NN output energies. A configuration was ultimately chosen for new DFT calculations if the energetic criterion was bigger than 0.1 eV.

After excluding the non-converged configurations in DFT calculations, a total of 2285, 2165, 2047 data points were used to fit the ground, neutral and anionic state EANN PESs, respectively, among which 90% of data points were used for training, while the remaining 10% were reserved for validation. These EANN hyperparameters take values of $L = 0$-$2$, $r_c = 7.6$ Å, $\alpha = 0.53$ Å$^{-2}$, and $r_s = 0.0$-$7.4$ Å every 0.6 Å, yielding 39 EAD features. The atomic neural network comprises two hidden layers with 30 and 40 neurons for the ground, neutral and anionic PESs. The overall root mean square errors (RMSEs) of the ground, neutral, and anionic PESs are 11.8, 59.9, and 75.8 meV for energies, and 17.7, 60.4, and 113.0 meV/Å for atomic forces, respectively. The significantly larger errors in the neutral and anionic state PESs compared to the ground state PES primarily stem from the broader energy range of diabatic states and extra errors in CDFT calculations.

**I.B2 The PW91-based PESs**

While the potential energy landscape has been shown to critically regulate adiabatic energy transfer dynamics(*11, 12*), its role in nonadiabatic energy transfer dynamics during molecule-metal scattering remains relatively unclear. Herein, we further construct diabatic state PESs using the PW91 functional, based on which Born-Oppenheimer molecular dynamics (BOMD) predicted a large amount of vibrational energy loss after NO scattering from Au(111)(*12, 13*). As will be discussed in Section Ⅱ.A, the energy differences between the neutral and ground states ($U_0$-$E_g$) and between the anionic and ground states ($U_1$-$E_g$) calculated using the revPBE and PW91 functionals are very similar. To avoid performing costly CDFT calculations for thousands of configurations required for PES fitting, we add ($U_0^{RE}$-$E_g^{RE}$) calculated on the revPBE PES on top of the PW91 ground state ($E_g^{PW}$) PES(*12, 13*) reported previously to construct the PW91 neutral state ($U_0^{PW}$) PES,

$$U_0^{PW} = E_g^{PW} + (U_0^{RE} - E_g^{RE}).  \tag{S8}$$

The PW91 anionic state PES can also be obtained by,

$$U_1^{PW} = E_g^{PW} + (U_1^{RE} - E_g^{RE}),  \tag{S9}$$

where ($U_1^{RE} - E_g^{RE}$) is the energy difference between the anionic and ground states calculated on the revPBE PES.

### I.C Independent electron surface hopping dynamics

Based on the full-dimensional ground and diabatic PESs at first-principles level, we applied the independent electron surface hopping (IESH) method to evolve coupled nuclear and electronic dynamics(*14, 15*). The method is based on a discretized version of the Newns–Anderson (NA) Hamiltonian(*16*), which describes an impurity level (here the lowest unoccupied molecular orbital (LUMO) of the molecule) interacting with a manifold of electronic states with the assumption that the metallic electrons are non-interacting. The many-electron News-Anderson Hamiltonian is written as,

$$\begin{aligned}H_{el}(\mathbf{R}) = &U_0(\mathbf{R}) + (U_1(\mathbf{R}) - U_0(\mathbf{R}))a^+ a + \int_{E_-}^{E_+} d\varepsilon (\varepsilon c_\varepsilon^+ c_\varepsilon) \\ &+ \int_{E_-}^{E_+} d\varepsilon (V(\varepsilon;\mathbf{R})a^+ c_\varepsilon + V(\varepsilon;\mathbf{R})^* c_\varepsilon^+ a)\end{aligned}.  \tag{S10}$$

Here, $U_0(\mathbf{R})$ is the PES for the interaction of a neutral molecule (with the LUMO orbital, $|a\rangle$, unfilled) with the metal surface, $U_1(\mathbf{R})$ is the PES for the interaction of a negatively charged molecule (with $|a\rangle$ filled) with the metal surface, $\hat{a}^+$ ($\hat{a}$) is the creation (annihilation) operator for an electron in $|a\rangle$. The third term in Eq. (S10) describes a continuum of non-interacting electrons, *i.e.*, the metallic band, where $\hat{c}_\varepsilon^+$ ($\hat{c}_\varepsilon$) is the creation (annihilation) operator for an electron in the metal band with energy $\varepsilon$. The last term couples the molecular orbital and the metal band, which we assume to be $\varepsilon$-independent (wide band limit). The practical IESH Hamiltonian is obtained by discretizing the continuous band using a Gauss–Legendre quadrature with unevenly distributed metal orbitals,(14)

$$H_{el}(\mathbf{R}) = U_0(\mathbf{R}) + (U_1(\mathbf{R}) - U_0(\mathbf{R}))a^+ a + \sum_{k=1}^{M} \varepsilon_k c_k^+ c_k + \sum_{k=1}^{M} V_{ak}(\mathbf{R})(a^+ c_k + c_k^+ a)$$ (S11)

Note that the metal orbital energy ($\varepsilon_k$) and the coupling ($V_{ak}$) between the molecular orbital ($|a\rangle$) and the metal orbital ($|k\rangle$) are defined by,

$$\varepsilon_k = \pm \frac{\Delta E}{2}\left(\frac{1}{2} + \frac{1}{2}x_k\right),$$ (S12)

$$V_{ak} = V_c \times \frac{\sqrt{2w_k}}{2},$$ (S13)

where $\Delta E$ is the bandwidth and $x_k/w_k$ are the knot points/weights for Gauss-Legendre quadrature in the interval [-1,1].

Due to the assumption of independent electrons, this Hamiltonian (Eq. (S11)) can be written as the sum of one-electron Hamiltonians, each of which is given by

$$H_{el}^1(\mathbf{R}) = (U_1(\mathbf{R}) - U_0(\mathbf{R}))|a\rangle\langle a| + \sum_{k=1}^{M} \varepsilon_k |k\rangle\langle k| + \sum_{k=1}^{M} V_{ak}(\mathbf{R})(|a\rangle\langle k| + |k\rangle\langle a|)$$ (S14)

By diagonalizing Eq. (S14), a series of eigenvalues ($E_j$) and eigenstates $|\varphi_j(\mathbf{R})\rangle$ are obtained. Thus, the adiabatic Hamiltonian for evolving the nuclear subsystem can be expressed with the eigenvalues,

$$H(\mathbf{R},\mathbf{P}) = \sum_{i=1}^{3N} \frac{P_i^2}{2M_i} + U_0(\mathbf{R}) + \sum_{j \in \mathbf{s}(t)} E_j(\mathbf{R}),$$ (S15)

where the total energy of electrons is represented by the sum of a set of occupied single electronic states determined by a time-dependent vector **s**(*t*). The corresponding $N_e$-electron eigenstate of the many-electron Hamiltonian can then be obtained by populating $N_e$ of these one-electron orbitals in a Slater determinant,

$$|\mathbf{j}\rangle = |\varphi_{j_1}\varphi_{j_2},\ldots,\varphi_{j_{N_e}}|. \tag{S16}$$

In this representation, the ground state corresponds to filling $N_e$ (half of the number of metal orbitals, *M*) electrons from the lowest single-electron orbital up to the Fermi level, while the excited states correspond to the determinants in which one or more electrons jump to empty orbitals above the Fermi level.

In IESH, the total many-electron wave function is a single Slater determinant state,

$$|\Psi\rangle = |\psi_1\psi_2,\ldots,\psi_{N_e}|. \tag{S17}$$

Each electron in the system is evolved independently according to the one-electron Hamiltonian $H_{el}^1$,

$$i\hbar \frac{\partial |\psi_\alpha\rangle}{\partial t} = H_{el}^1(\mathbf{R})|\psi_\alpha\rangle, \tag{S18}$$

where $\alpha$ goes from 1 to $N_e$. We expand each single-electron wavefunction in terms of ($M$ + 1) electronic basis functions,

$$\psi_\alpha(\mathbf{r},\mathbf{R},t) = \sum_j c_j^\alpha(t)\varphi_j(\mathbf{r},\mathbf{R}), \tag{S19}$$

where $\{c_j^\alpha\}$ are the complex expansion coefficients for $\alpha$th electron. Substituting Eq. (S19) into Eq. (S18), we get,

S10

$$i\hbar \dot{c}_k^\alpha = E_k(\mathbf{R})c_k^\alpha - i\hbar \sum_{j \neq k} \dot{\mathbf{R}} \cdot \mathbf{d}_{kj}(\mathbf{R})c_j^\alpha, \qquad (S20)$$

where $\mathbf{d}_{kj}$ is the nonadiabatic coupling between orbitals $k$ and $j$.

In practice, Eqs. (S15) and (S20) describe the evolution of the nuclear and electronic subsystems. Within the IESH approach, the hopping probability between electronic states represented by $|\mathbf{k}\rangle$ and $|\mathbf{j}\rangle$ are non-zero only when $|\mathbf{k}\rangle$ and $|\mathbf{j}\rangle$ differ in exactly one occupied orbital, that is, $k_1=j_1$, $k_2=j_2$, …, $k_{N_e}=j_{N_e}$, except for a single orbital at which they differ ($k_i \neq j_i$). And the probability of a hop from the occupied orbital $k$ to an unoccupied orbital $j$ is calculated by,

$$g_{k \to j} = \max\left\{ \frac{-2\,\text{Re}(A_{kj})}{A_{kk}} \dot{\mathbf{R}} \cdot \mathbf{d}_{jk} \Delta t, 0 \right\}, \qquad (S21)$$

where $\Delta t$ is the electronic time step, $A_{kj}$ and $A_{kk}$ are elements of electronic density matrix,

$$A_{kj} = \langle \mathbf{k}|\psi\rangle\langle\psi|\mathbf{j}\rangle. \qquad (S22)$$

Each inner product in Eq. (S20) is calculated from the overlap matrix $\mathbf{S}$,

$$\langle \mathbf{k}|\Psi\rangle = |\mathbf{S}|, \qquad (S23)$$

with the elements of $\mathbf{S}$ given by,

$$S_{\alpha\beta} = c_{k_\alpha}^\beta, \qquad (S24)$$

where $k_\alpha$ represents the orbital occupied by the $\alpha$th electron in the eigenstate $|\mathbf{k}\rangle$, $\mathbf{c}^\beta$ is the wavefunction expansion coefficient of electron $\beta$. After these modifications, the IESH method follows the conventional fewest-switch surface hopping (FSSH) algorithm(*14, 17*).

In this work, the initial electronic occupation at a given electronic temperature ($T_e$) is sampled using the following procedure(*18*). First, the $N_e$ lowest metal orbitals are occupied. Then, a hop attempt is made from an occupied orbital *i* to an unoccupied orbital *a*, with *i* and *a* selected randomly. If $\varepsilon_a<\varepsilon_i$, the hop attempt is accepted, *i.e.*, orbital *i* becomes unoccupied and orbital *a* becomes occupied. If $\varepsilon_a>\varepsilon_i$, the hop attempt is accepted with a probability of $\exp[-(\varepsilon_a-\varepsilon_i)/(k_bT_e)]$. The electronic occupation obtained after 50000 times hop attempts is chosen as the initial electronic occupation before a trajectory propagation.

**I.D. Quasi-classical trajectory implementation**

To compare with scattering experiments, we have performed quasi-classical trajectory (QCT) calculations for the nuclear dynamics. The Jacobi coordinate of the NO/Au(111) system for describing the state-to-state scattering process is illustrated in Fig. S1. The BOMD and IESH simulations were performed using an in-house modified version of VENUS code(*19, 20*). In this work, all trajectories started with initially nonrotating NO molecules positioned with the center of mass 8.0 Å above the topmost layer of Au(111), *i.e.*, $Z_{NO}$ = 8.0 Å (see Fig. S1). The lateral coordinates of the center of mass of NO (*X* and *Y*) were randomly sampled in the simulation supercell. The initial polar (*θ*) and azimuthal (*φ*) angles of the molecular internuclear vector (defined as pointing from N to O) were randomly sampled unless stated otherwise. The initial values of the internuclear distance, $r_{NO}$, and the conjugated momentum were sampled semi-classically for given vibrational and rotational quantum numbers $v$ and $j$(21). The initial condition of surface atoms was chosen from the Anderson thermostat, but no thermostat was imposed during the collisional process. Nuclear coordinates and momenta were propagated using the Verlet algorithm with a time step of 0.1 fs, with a maximum simulation time of 10 ps. The trajectory was terminated and assigned to a "scattered" event when molecule-surface vertical distance exceeded 8.0 Å and the molecule velocity pointed away from the surface, for which the final vibrational state ($v_f$) of NO was determined by the Einstein−Brillouin−Keller (EBK) quantization(22) and rotational quantum number ($j_f$) was obtained from the quantum mechanical expression for rotational angular momentum. Alternatively, the trajectory was labelled as "trapped" if the trajectory exceeded the maximum simulation time. To obtain the final rovibrational state distributions, the fractional vibrational and rotational quantum numbers were binned into the nearest integers via a histogram binning procedure.

## II. Supplementary Results

**II. A Validation of the construction of PW91 diabatic PESs**

In Fig. S2, we compared ground and diabatic state energies as well as the energy differences between the neutral and ground states ($U_0$-$E_g$) and between the anionic and ground states ($U_1$-$E_g$) computed by the revPBE and PW91 functionals. Although the ground and diabatic state energies have discrepancies, the ($U_0$-$E_g$) and ($U_1$-$E_g$) calculated by the two functionals are quite similar. Quantitatively, the root-mean-square deviations of ($U_0$-$E_g$) and ($U_1$-$E_g$) calculated by the two functionals in Fig. S2 are 23.3 meV and 34.0 meV, respectively—both less than half of the fitting RMSEs of the revPBE neutral and anionic PESs, validating our strategy to construct the PW91 diabatic PESs.

**II. B Convergence tests of IESH dynamics**

Ground state energy calculated from the NA Hamiltonian can be written as,

$$H(\mathbf{R},\mathbf{P}) = U_0(\mathbf{R}) + \sum_{j=1}^{N_e} E_j(\mathbf{R}), \tag{S25}$$

where $\{E_j\}$ are the eigenvalues of the NA Hamiltonian as mentioned above. In principle, NA ground state energy should match the true ground state energy calculated from DFT. We ensure this by multiplying a rescaling factor $S(\mathbf{R})$ on the coupling between molecular and metal orbitals defined in Eq. (S13),

$$V_{ak}(\mathbf{R}) = S(\mathbf{R}) \times V_c(\mathbf{R}) \times \frac{\sqrt{2w_k}}{2}, \tag{S26}$$

where $V_c(\mathbf{R})$ is the diabatic coupling. The rescaling factor is coordinate-dependent and is fitted using the NN method, similar to the training of the PES. For the NO/Au(111) system, $S(\mathbf{R})$ ranges approximately from 0.8 to 1.3 across the entire configuration space, indicating that it serves as only a minor correction to the diabatic coupling $V_c(\mathbf{R})$. After this adjustment, the $E_g$ calculated from the NA Hamiltonian aligns well with that obtained from the ground state PES, as shown in Fig. S3. Additionally, BOMD results simulated based on the ground state PES and the lowest state of the NA Hamiltonian also agree well with each other, further validating the rescaling algorithm.

We also tested the convergence of the number of metal states $M$ in Fig. S4, ultimately selecting $M = 40$ to obtain converged results with an affordable computational cost. It is also shown that the final results are not very sensitive to the bandwidth $\Delta E$, and we finally chose $\Delta E = 7.0$ eV, consistent with Tully's work(*14, 23*).

## II. C Results from all trajectories

The state-to-state scattering results reported by Wodtke and coworkers do not include final states with $v_f = 0$, $v_f > v_i$ or $v_f = 1$ (when $v_i = 11$ and 16), due to practical considerations and scope limitations. Consequently, these final states are excluded from our analysis in Fig. 1. Additionally, the final state populations in the main text were derived solely from single-bounce scattering trajectories. For completeness, the scattering distributions in Figs. 1 and 3 of the main text are reproduced here in Figs. S5 and S7 with all computed final states included and all bounce events accounted for. On the revPBE and PW91 PESs, the results remain largely unchanged, while multibounce trajectories significantly affect the final state populations on the RST PES, especially for the $E_i$-dependence of vibrational inelasticity depicted in Fig. S7.

**Table S1.** Energies and geometries of stationary points of NO on Au(111) optimized by the revPBE and PW91 functionals. $\theta$ refers to the angle between the N-O vector and the surface normal. Values in the parentheses are the results optimized on the corresponding ground state EANN PESs.

| Functional | Stationary point | Energy (eV) | $r_{NO}$ (Å) | $Z_{NO}$ (Å) | $\theta$ (deg) |
|---|---|---|---|---|---|
| revPBE | Adsorption | -0.14 (-0.15) | 1.17 (1.17) | 2.85 (2.87) | 62 (56) |
|  | TS | 3.58 (3.55) | 2.11 (2.09) | 1.09 (1.04) | 77 (78) |
|  | Product | 2.63 (2.62) | 3.18 (3.20) | 0.83 (0.83) | 89 (89) |
| PW91 | Adsorption | -0.36 (-0.39) | 1.17 (1.17) | 2.75 (2.76) | 57 (54) |
|  | TS | 2.88 (2.86) | 1.88 (1.89) | 1.28 (1.27) | 89 (88) |
|  | Product | 2.41 (2.40) | 2.57 (2.56) | 1.15 (1.15) | 88 (89) |

**Table S2.** Average vibrational ($\Delta E_{vib}$), rotational ($\Delta E_{rot}$), translational ($\Delta E_{trans}$), surface phonon ($\Delta E_{ph}$) and surface electron ($\Delta E_{el}$) energy changes (in eV) of NO scattering from Au(111) in BOMD and IESH simulations based on the revPBE PES.

| initial state | method | $\Delta E_{vib}$ | $\Delta E_{trans}$ | $\Delta E_{rot}$ | $\Delta E_{ph}$ | $\Delta E_{el}$ |
|---|---|---|---|---|---|---|
| $v_i$=3 | BOMD | -0.020 | -0.661 | 0.252 | 0.429 | 0.000 |
| $E_i$=1.08 eV | IESH | -0.167 | -0.650 | 0.257 | 0.427 | 0.133 |
| $v_i$=11 | BOMD | -0.462 | -0.430 | 0.314 | 0.578 | 0.000 |
| $E_i$=0.95 eV | IESH | -1.137 | -0.441 | 0.327 | 0.563 | 0.688 |
| $v_i$=16 | BOMD | -1.022 | -0.024 | 0.374 | 0.672 | 0.000 |
| $E_i$=0.52 eV | IESH | -1.866 | -0.074 | 0.326 | 0.556 | 1.058 |

**Table S3.** Same as Table S2, but on the PW91 PES.

| initial state | method | $\Delta E_{vib}$ | $\Delta E_{trans}$ | $\Delta E_{rot}$ | $\Delta E_{ph}$ | $\Delta E_{el}$ |
|---|---|---|---|---|---|---|
| $v_i$=3 | BOMD | -0.003 | -0.475 | 0.152 | 0.326 | 0.000 |
| $E_i$=1.08 eV | IESH | -0.112 | -0.468 | 0.151 | 0.330 | 0.099 |
| $v_i$=11 | BOMD | -0.077 | -0.476 | 0.179 | 0.374 | 0.000 |
| $E_i$=0.95 eV | IESH | -0.684 | -0.404 | 0.179 | 0.374 | 0.535 |
| $v_i$=16 | BOMD | -0.156 | -0.255 | 0.134 | 0.277 | 0.000 |
| $E_i$=0.52 eV | IESH | -1.129 | -0.178 | 0.134 | 0.274 | 0.900 |

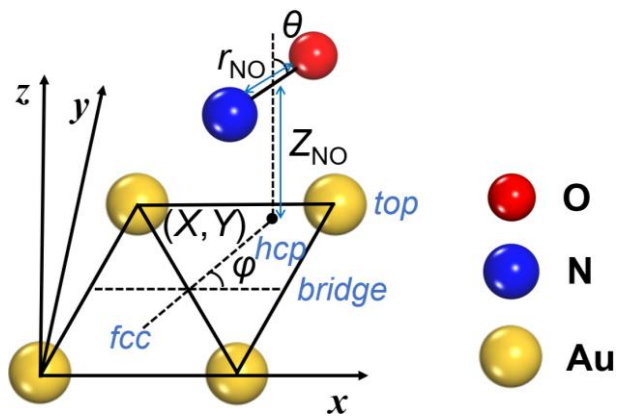

**Fig. S1. Jacobi coordinates for describing the NO/Au(111) system.** The N-O bond length ($r_{NO}$), the distance of the molecular center of mass to surface ($Z_{NO}$), the lateral coordinates of the molecular center of mass ($X$, $Y$), the polar angle ($\theta$) and azimuthal angle ($\varphi$) are labelled. $\theta = 0°$ corresponds to NO being perpendicular to the surface panel with the N atom facing the surface.

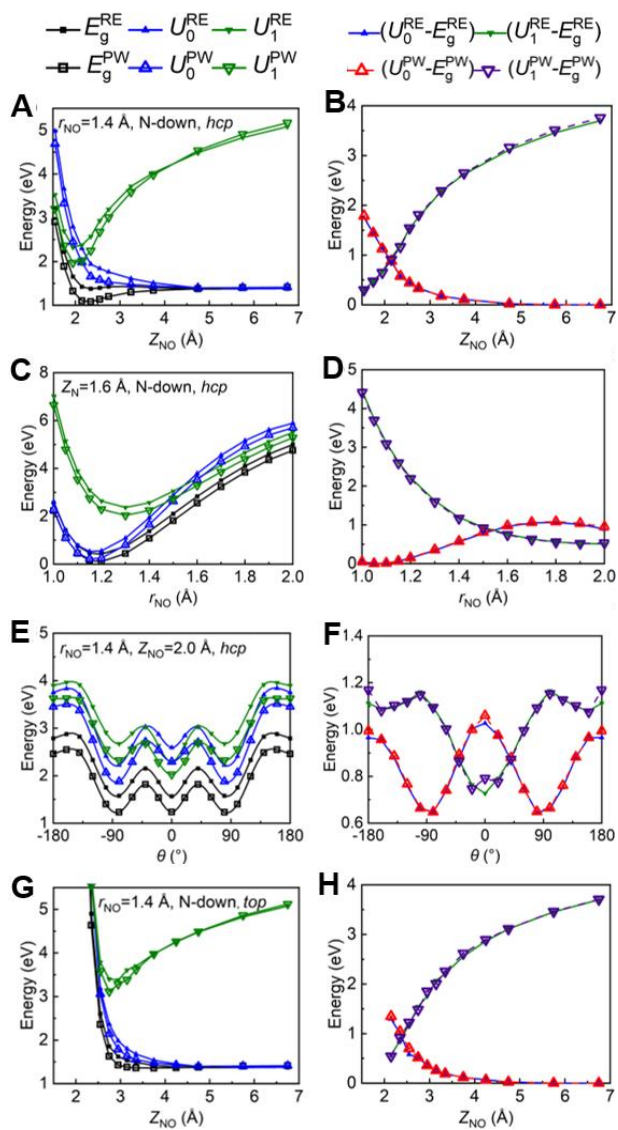

**Fig. S2. Comparison of ground and diabatic states calculated by the revPBE (labelled with 'RE', solid symbol) and PW91 (labelled with 'PW', hollow symbol) functionals.** (**A**), (**C**), (**E**) and (**G**) display the ground ($E_g$), neutral ($U_0$) and anionic ($U_1$) state energies. (**B**), (**D**), (**E**) and (**F**) compare the energy differences ($U_0$-$E_g$) and ($U_1$-$E_g$), calculated by revPBE and PW91 functionals. The NO molecule is placed on the hcp site of an optimized Au(111) surface. (**A**)-(**B**) $Z_{NO}$ is varied with $r_{NO}$ fixed at 1.4 Å and $\theta$ fixed at 0°. (**C**)-(**D**) $r_{NO}$ is varied with the height of N atom fixed at 1.6 Å. (**E**)-(**F**) $\theta$ is varied with $Z_{NO}$ fixed at 2.0 Å. (**G**)-(**H**) Same as (**A**)-(**B**), but on the top site. The energy is relative to $E_g$ of a free NO molecule far from the surface.

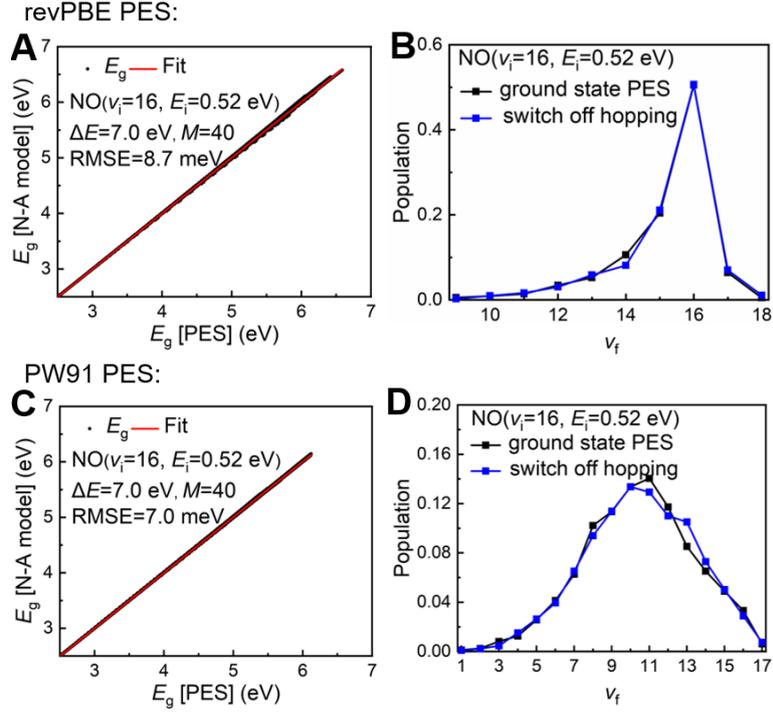

**Fig. S3. Validation of the rescaling algorithm.** (**A**) Comparison of ground state energies $E_g$ on the revPBE PES and calculated from the NA Hamiltonian matrix with $\Delta E = 7.0$ eV and $M = 40$. The used configurations are collected from a trajectory of NO($v_i = 16$, $E_i = 0.52$ eV) scattering from Au(111), and the RMSE of the two $E_g$ is 8.7 meV. (**B**) A comparison of BOMD results obtained using two methods is presented: one method directly simulates the system on the ground state PES, while the other uses IESH dynamics with the hopping between different electronic states disabled. (**C**)-(**D**) Similar to (**A**)-(**B**), but on the PW91 PES.

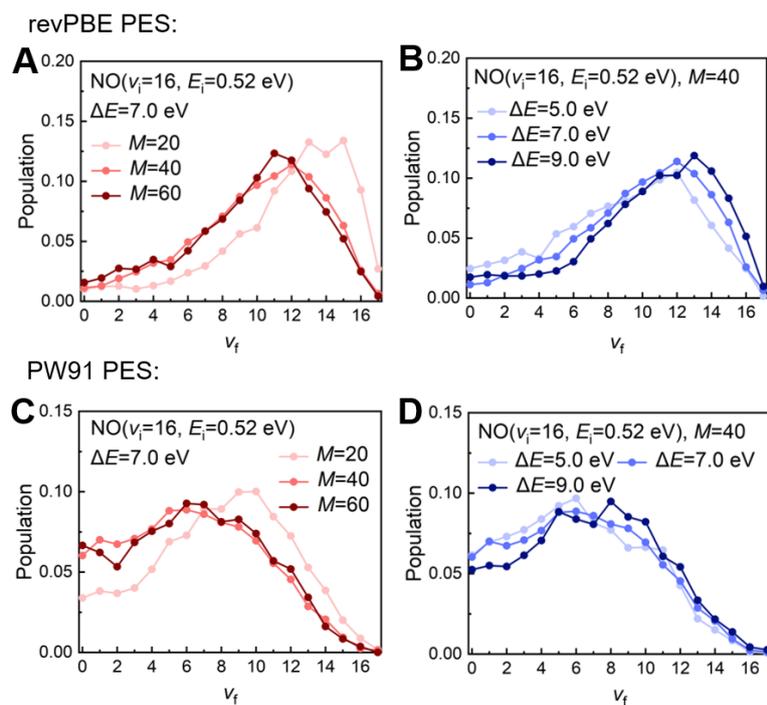

**Fig. S4. Tests of ΔE and M.** Final vibrational state distributions of NO($v_i$ = 16, $E_i$ = 0.52 eV) scattering from Au(111) with $M$ = 20, 40, 60 and ΔE = 7.0 eV, on the (**A**) revPBE and (**C**) PW91 PESs. Final vibrational state distributions of NO($v_i$ = 16, $E_i$ = 0.52 eV) scattering from Au(111) using different ΔE and $M$ = 40 on the (**B**) revPBE and (**D**) PW91 PESs.

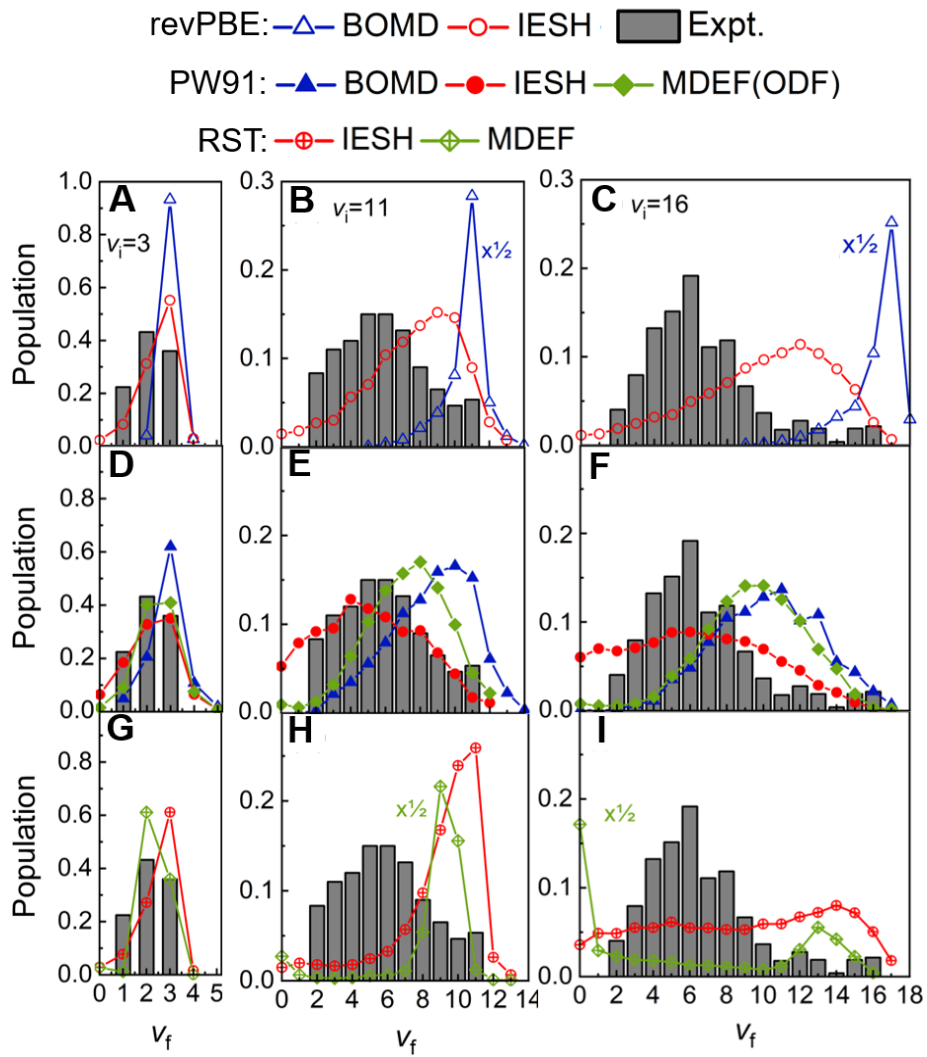

**Fig. S5.** Similar to Fig. 1, but trajectories from all final vibrational states and all bounce events are included.

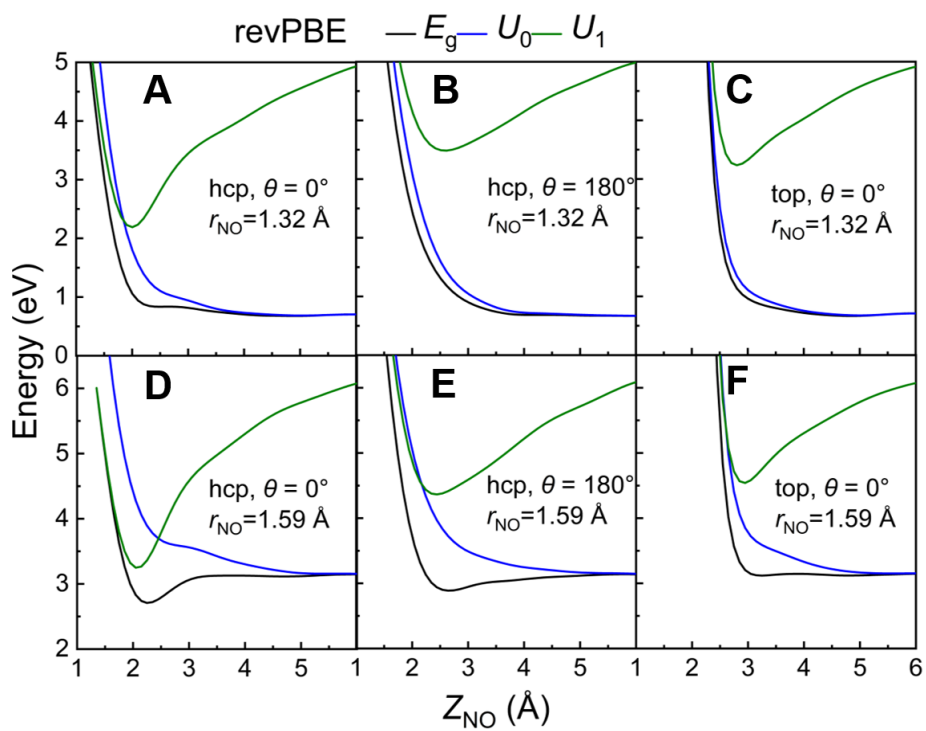

**Fig. S6. One dimensional cuts of the revPBE PES.** (**A**) One dimensional energy curves of the ground ($E_g$), neutral ($U_0$) and anionic ($U_1$) states on the revPBE PES, as a function of $Z_{NO}$, with $r_{NO}$ = 1.32 Å (approximately the classical outer turning point of NO($v$ = 3)) and $\theta$ = 0°. (**B**) Similar to (**A**), but with $\theta$ = 180°. (**C**) Similar to (**A**), but with NO placed on the top site. (**D**)-(**F**) Similar to (**A**)-(**C**), but with $r_{NO}$ fixed at 1.59 Å (approximately the classical outer turning point of NO($v$ = 16)). The energy is relative to the $E_g$ of a free NO molecule far from the surface.

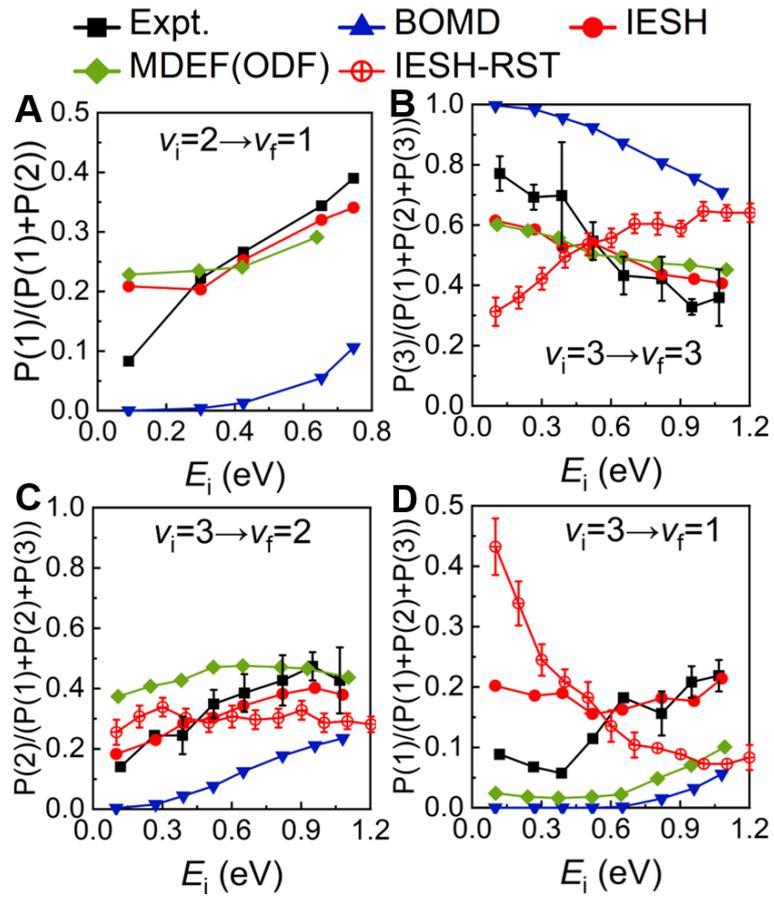

**Fig. S7.** Similar to Fig. 3, but trajectories with multi-bounce are also included.

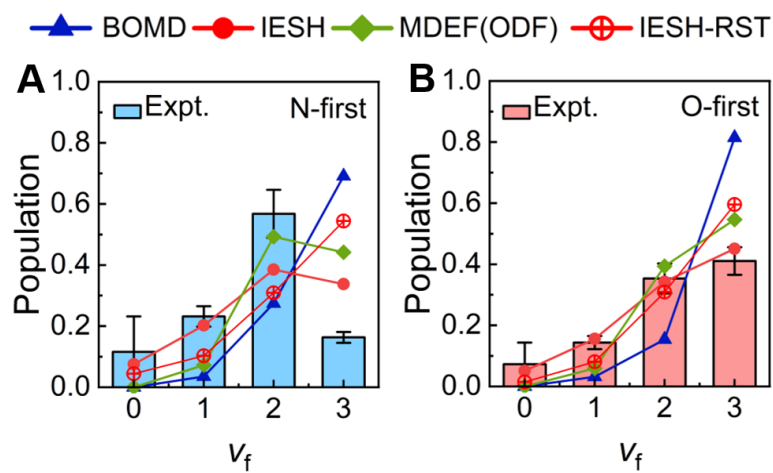

**Fig. S8. Dependence of final vibrational state distributions on initial orientation.** Experimental final vibrational state distributions(*24*) of NO ($v_i$ = 3, $E_i$ = 0.96 eV) scattering from Au(111) with (**A**) N-first and (**B**) O-first orientations, in comparison with BOMD, IESH and MDEF(ODF)(*25*) on the PW91 PES and IESH using the RST PES(*26*). Note that all theoretical models consider only single-bounce trajectories, and trajectories with final vibrational states beyond the experimental range are excluded.

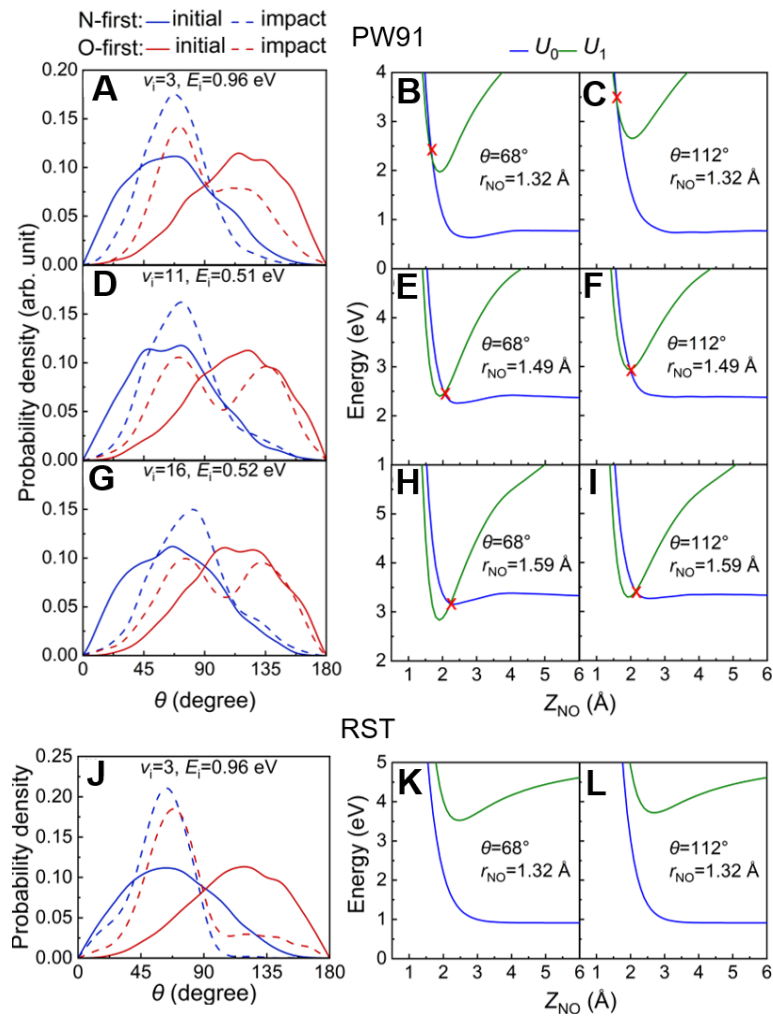

**Fig. S9. Polar angle distributions and one dimensional cuts on the PW91 and RST PESs.** Polar angle distributions at the initial and impact points for (**A**) $v_i$ = 3, $E_i$ = 0.96 eV, (**D**) $v_i$ = 11, $E_i$ = 0.51 eV and (**G**) $v_i$ = 16, $E_i$ = 0.52 eV, obtained on the PW91 PES. One dimensional energy curves of the neutral ($U_0$) and anionic ($U_1$) states on the PW91 PES, as a function of $Z_{NO}$, with $r_{NO}$ = 1.32 Å, $\theta$ = (**B**) 68° and (**C**) 112°. The NO molecule is placed on the hcp site of Au(111). (**E**)-(**F**) and (**H**)-(**I**) are analogous to (**B**)-(**C**), with $r_{NO}$ fixed at 1.49 Å and 1.59 Å, respectively. (**J**)-(**L**) are analogous to (**A**)-(**C**), but are presented on the RST PES. The energy is relative to the $E_g$ of a free NO molecule far from the surface. 'X' labels show crossing points of the neutral and anionic states.